\documentclass[12pt]{article}
\usepackage{a4}
\usepackage{amsfonts}
\usepackage{amsmath,amssymb}
\usepackage{graphicx}

\usepackage{multirow}
\usepackage[margin=0.8in]{geometry}

\newcommand{\overbar}[1]{\mkern1.5mu\overline{\mkern-1.5mu#1\mkern-1.5mu}\mkern 1.5mu}

\makeatother

\def\un#1{\relax\ifmmode\@@underline#1\else
        $\@@underline{\hbox{#1}}$\relax\fi}

\let\du=\du                     % dot-under
\def\bo{{\raise-.3ex\hbox{\large$\Box$}}}               % D'Alembertian
\def\TH{{\raise.2ex\hbox{$\displaystyle \bigodot$}\mskip-4.7mu \llap H \;}}
\def\face{{\raise.2ex\hbox{$\displaystyle \bigodot$}\mskip-2.2mu \llap {$\ddot
        \smile$}}}                                      % happy face
\def\leftrightarrowfill{$\mathsurround=0pt \mathord\leftarrow \mkern-6mu
        \cleaders\hbox{$\mkern-2mu \mathord- \mkern-2mu$}\hfill
        \mkern-6mu \mathord\rightarrow$}
\def\dvec#1{\vbox{\ialign{##\crcr
        \leftrightarrowfill\crcr\noalign{\kern-1pt\nointerlineskip}
        $\hfil\displaystyle{#1}\hfil$\crcr}}}           % <--> accent
\def\frac#1#2{{\textstyle{#1\over\vphantom2\smash{\raise.20ex
        \hbox{$\scriptstyle{#2}$}}}}}                   % fraction
\def\sfrac#1#2{{\vphantom1\smash{\lower.5ex\hbox{\small$#1$}}\over
        \vphantom1\smash{\raise.4ex\hbox{\small$#2$}}}} % alternate fraction
\def\bfrac#1#2{{\vphantom1\smash{\lower.5ex\hbox{$#1$}}\over
        \vphantom1\smash{\raise.3ex\hbox{$#2$}}}}       % "
\def\afrac#1#2{{\vphantom1\smash{\lower.5ex\hbox{$#1$}}\over#2}}    % "
\def\[{\lfloor{\hskip 0.35pt}\!\!\!\lceil}
\def\]{\rfloor{\hskip 0.35pt}\!\!\!\rceil}

\def\du#1#2{_{#1}{}^{#2}}

\def\un{\underline}
\def\fracmm#1#2{{{#1}\over{#2}}}

\def\low#1{{\raise -3pt\hbox{${\hskip 0.75pt}\!_{#1}$}}}

\newskip\humongous \humongous=0pt plus 1000pt minus 1000pt

\newif\ifdtup

\newcommand{\be}{\begin{equation}}
\newcommand{\ee}{\end{equation}}
\newcommand{\nbe}{\begin{equation*}}
\newcommand{\nee}{\end{equation*}}

\begin{document}

\thispagestyle{empty}

{\hbox to\hsize{
\vbox{\noindent revised 2017 \hfill IPMU16-0098 }}}

\noindent
\vskip2.0cm
\begin{center}

{\Large\bf SUSY breaking after inflation in supergravity \\
\vglue.1in
with inflaton in a massive vector supermultiplet}

\vglue.3in

Yermek Aldabergenov~${}^{a}$  and Sergei V. Ketov~${}^{a,b,c,d}$ 
\vglue.1in

${}^a$~Department of Physics, Tokyo Metropolitan University, \\
Minami-ohsawa 1-1, Hachioji-shi, Tokyo 192-0397, Japan \\
${}^b$~Kavli Institute for the Physics and Mathematics of the Universe (IPMU),
\\The University of Tokyo, Chiba 277-8568, Japan \\
${}^c$~Department of Physics, Faculty of Science, Chulalongkorn University,\\
Thanon Phayathai, Pathumwan, Bangkok 10330, Thailand\\
${}^d$~Institute of Physics and Technology, Tomsk Polytechnic University,\\
30 Lenin Ave., Tomsk 634050, Russian Federation \\
\vglue.1in
aldabergenov-yermek@ed.tmu.ac.jp, ketov@tmu.ac.jp
\end{center}

\vglue.3in

\begin{center}
{\Large\bf Abstract}
\end{center}
\vglue.1in
\noindent  We propose a limited class of models, describing interacting chiral multiplets with a non-minimal coupling to a vector multiplet, in curved superspace of $N=1$ supergravity. Those models are suitable for the inflationary model building in supergravity with inflaton assigned to a massive vector multiplet and spontaneous SUSY breaking in Minkowski vacuum after inflation, for any values of the inflationary parameters $n_s$ and $r$, and any scale of SUSY
breaking.

\newpage

\section{Introduction}

Success of the inflationary scenario for early Universe is, on the one hand,  due to overcoming the theoretical problems (horizon, flatness, structure formation) of the standard (Einstein-Friedmann) cosmology and, on the other hand, due to its remarkable agreement with the CMB observational data (COBE, WMAP, PLANCK). For instance, the observed breaking of CMB scale invariance  is measured by the scalar tilt, $n_{\text{s}}-0.9666=\pm 0.0062$~\cite{Ade:2015xua, Ade:2015lrj}, and the relative magnitude of primordial gravity waves is parametrized by the tensor-to-scalar ratio  $r<0.07$~\cite{Array:2015xqh}. Those observations favour chaotic slow-roll inflation in its single-field realization,  i.e. the large-field inflation driven by a single scalar called {\it inflaton} with an approximately flat scalar potential.

Embedding a single-field inflation into $N=1$ four-dimensional supergravity is needed to connect inflationary models to particle physics beyond the Standard Model, and towards their ultimate embedding into string theory. It requires inflaton to belong to a massive $N=1$ multiplet that can be either a chiral multiplet (of the highest spin 1/2) or a real vector multiplet (of the highest spin 1).  Most of the literature about inflation in supergravity uses the first option --- see e.g., the reviews \cite{Yamaguchi:2011kg,Ketov:2012yz}  --- since it is usually assumed that vector fields do not play any role during inflation.~\footnote{Taking into account vector fields is believed to be important after inflation, during reheating.} However, assuming inflaton to be in a chiral multiplet also causes some problems. First, the scalar component of a chiral multiplet is {\it complex}, which implies the need to stabilize another (non-inflaton) scalar during inflation. Second, there is also the so-called $\eta$-{\it problem} caused by the presence of the exponential factor $e^K$ in the scalar potential of supergravity with chiral superfields, which generically prevents slow roll. Third, there are problems also with ensuring the inflaton scalar potential to be bounded from below, and with getting SUSY breaking in a Minkowski vacuum after inflation too. Of course, the inflationary model building in supergravity now has many models that overcome some of those problems --- see e.g., \cite{Goncharov:1983mw,Kawasaki:2000yn,Kallosh:2010ug,Kallosh:2010xz,Abe:2014opa} and references therein. However, it often comes at the price of having more matter superfields together with a need to invent the dynamics for them.  The {\it minimal} inflationary models with a single inflaton chiral superfield, with or without SUSY breaking after inflation, are also possible 
\cite{Ketov:2014qha,Ketov:2014hya,Ketov:2016gej} but require tuning both K\"ahler potential and a superpotential. 
Yet another approach, based on the use of non-linear realizations of SUSY and nilpotent chiral superfields, was introduced to the supergravity-based inflationary model building in \cite{Ferrara:2014kva}.

When inflaton is assigned to a massive {\it vector} multiplet, there is no need of its complexification, because the scalar field component of a real massive $N=1$ vector multiplet is {\it real}. Accordingly, there is no need for other scalars and their stabilization during inflation, in the minimal supergravity setup. The $\eta$-problem also does not arise because the scalar potential of a vector multiplet in supergravity has a different structure (of the $D$-type instead of the $F$-type). Actually, the corresponding minimal inflationary models were already constructed by Ferrara, Kallosh, Linde and Porrati in \cite{Ferrara:2013rsa} by exploiting the non-minimal self-coupling of a vector multiplet to supergravity, found by Van Proeyen  in \cite{VanProeyen:1979ks}.

The supergravity inflationary models of  \cite{Ferrara:2013rsa} have the single-field scalar potential given by an
arbitrary real function squared. Those scalar potentials are always bounded from below and allow any
desired values of $n_s$ and $r$. However, the minima of the scalar potentials of \cite{Ferrara:2013rsa} 
have the vanishing cosmological constant and the vanishing VEV of the auxiliary field $D$, so that they only have Minkowski vacua where supersymmetry is always restored after inflation. It is desirable to have more theoretical flexibility, as regards SUSY breaking, for phenomenological purposes.

In this paper we propose a simple extension of the inflationary models \cite{Ferrara:2013rsa} by adding a {\it Polonyi} (chiral) superfield \cite{Polonyi:1977pj}. Our models  also can accommodate arbitrary values of  $n_s$ and $r$, but have a Minkowski vacuum after inflation, with spontaneously broken supersymmetry (SUSY).

  Our paper is organized as follows. In Sec.~2 we propose a new class of supergravity models in curved superspace of $N=1$ old-minimal supergravity. Our models can be considered as the extensions of those  in \cite{Ferrara:2013rsa} via adding a chiral superfield and its coupling to a vector (inflaton) superfield in supergravity. We also compute the bosonic kinetic terms and the scalar potential in our models. In Sect.~3 we identify the chiral sector with the Polonyi model, and find a Minkowski vacuum with spontaneously broken supersymmetry after inflation that is not affected by the Polonyi superfield. Sect.~4 is our Conclusion.

\section{A vector multiplet non-minimally coupled to a chiral multiplet in supergravity}

Let us consider some chiral superfields $\Phi_i$ with arbitrary K{\"a}hler potential $K=K(\Phi_i,\overbar{\Phi}_i)$ and a chiral superpotential $\mathcal{W}=\mathcal{W}(\Phi_i)$, interacting with a real superfield  $V$ whose arbitrary potential is described by a real function $J=J(V)$. The real vector superfield $V$ is supposed to describe a massive
vector multiplet, while the chiral superfields  are supposed to be (gauge) singlets in our construction.

We employ the curved superspace formalism of $N=1$ supergravity \cite{Wess:1992cp}. Our notation and conventions coincide with the standard ones in \cite{Wess:1992cp}, including the spacetime signature 
$(-,+,+,+)$.~\footnote{The $N=1$ superconformal calculus used in \cite{Ferrara:2013rsa,VanProeyen:1979ks}
is equivalent to the curved superspace description \cite{Wess:1992cp} of $N=1$ Poincar\'e supergravity after the superconformal gauge fixing.}

Our models are defined by the Lagrangian ($M_{\rm Pl}=1$)
\begin{equation} \label{sslag}
\mathcal{L}=\int d^2\theta 2\mathcal{E}\left\lbrace \frac{3}{8}(\overbar{\mathcal{D}}\overbar{\mathcal{D}}-8\mathcal{R})e^{-\frac{1}{3}(K+2J)}+\frac{1}{4}W^\alpha W_\alpha +\mathcal{W} \right\rbrace +{\rm h.c.}~,
\end{equation}
where we have introduced the chiral density superfield $2\mathcal{E}$, the chiral scalar curvature superfield 
$\mathcal{R}$, and the chiral vector superfield strength $W_\alpha\equiv-\frac{1}{4}(\overbar{\mathcal{D}}\overbar{\mathcal{D}}-8\mathcal{R})\mathcal{D}_\alpha V$.

In order to calculate the bosonic part of our models, we set all fermions to zero, and define the bosonic field components of the relevant superfields. As regards the supergravity multiplet, we have
\begin{gather*}
2\mathcal{E}|=e,\;\;\;\;\mathcal{D}\mathcal{D}(2\mathcal{E})|=4e\overbar{M}~,\\
\mathcal{R}|=-\frac{1}{6}M,\;\;\;\;\mathcal{D}\mathcal{D}\mathcal{R}|=-\frac{1}{3}R+\frac{4}{9}M\overbar{M}+\frac{2}{9}b_mb^m-\frac{2}{3}i\mathcal{D}_mb^m~,
\end{gather*}
where we have introduced the vierbein determinant $e\equiv\text{det} e_m^a$, the spacetime scalar curvature
$R$, and the old-minimal set of the supergravity auxiliary fields, the complex scalar $M$ and the real vector $b_m$. 
The vertical bars denote the leading field components of a superfield at $\theta=\bar{\theta}=0$. 

The field components of $\Phi_i$ and $V$ are defined by
\begin{gather*}
\Phi_i|=A_i~,\;\;\;\;\mathcal{D}_\alpha\mathcal{D}_\beta\Phi_i|=-2\varepsilon_{\alpha\beta} F_i~,\;\;\;\;\overbar{\mathcal{D}}_{\dot{\alpha}}\mathcal{D}_{\alpha}\Phi_i|=-2i{\sigma_{\alpha\dot{\alpha}}}^m\partial_mA_i~,
\\\overbar{\mathcal{D}}\overbar{\mathcal{D}}\mathcal{D}\mathcal{D}\Phi_i|=16\Box A_i+\frac{32}{3}ib_a\partial^aA_i+\frac{32}{3}F_iM~,
\\V|=C~,\;\;\;\;\mathcal{D}_\alpha\mathcal{D}_\beta V|=\varepsilon_{\alpha\beta} X~,\;\;\;\;\overbar{\mathcal{D}}_{\dot{\alpha}}\mathcal{D}_{\alpha}V|={\sigma_{\alpha\dot{\alpha}}}^m(B_m-i\partial_mC)~,
\\\mathcal{D}_\alpha W^{\beta}|\equiv -\frac{1}{4}\mathcal{D}_\alpha(\overbar{\mathcal{D}}\overbar{\mathcal{D}}-8\mathcal{R})\mathcal{D}^{\beta}V=\frac{1}{2}{\sigma_{\alpha\dot{\alpha}}}^m\overbar{\sigma}^{\dot{\alpha}\beta n}(\mathcal{D}_m\partial_n C+iF_{mn})+{\delta_\alpha}^\beta (D+\frac{1}{2}\Box C)~,
\\\overbar{\mathcal{D}}\overbar{\mathcal{D}}\mathcal{D}\mathcal{D}V|=\frac{16}{3}b^m(B_m-i\partial_mC)+8\Box C-\frac{16}{3}MX+8D~,
\end{gather*}
in terms of the physical fields $A_i$, $C$, $B_m$ as complex scalars, a real scalar, and a real vector respectively, the chiral auxiliary fields $F_i$ and $X$ as complex scalars, the real auxiliary field $D$ as
a real scalar, and the vector field strength $F_{mn}=\mathcal{D}_mB_n-\mathcal{D}_nB_m$ of $B_m$. 

Using those definitions, we find by a straightforward calculation that the kinetic part of our Lagrangian is given by
\begin{multline}
e^{-1}\mathcal{L}_{\rm kin.}=e^{-\frac{1}{3}(K+2J)}\biggl\{ -\frac{1}{2}R-K_{ij*}\partial_m A_i\partial^m\bar{A}_j-\frac{1}{6}K_iK_j\partial_m A_i\partial^m A_j-\frac{1}{6}K_{i^*}K_{j^*}\partial_m \bar{A}_i\partial^m\bar{A}_j\\-\bigg(\frac{1}{3}{J'}^2-\frac{1}{2}J''\bigg)\partial_m C\partial^m C +\bigg(\frac{1}{3}{J'}^2-\frac{1}{2}J''\bigg)B_m B^m+J'\Box C +\frac{i}{3}J'B_m(K_{i^*}\partial^m\bar{A}_i-K_i\partial^m A_i)\\- \frac{1}{3}J'\partial_m C(K_{i^*}\partial^m\bar{A}_i+K_i\partial^m A_i) \biggl\}-\frac{1}{4}F_{mn}F^{mn}~,
\end{multline}
while its auxiliary part reads
\begin{multline}
e^{-1}\mathcal{L}_{\rm aux.}=e^{-\frac{1}{3}(K+2J)}\biggl\{\frac{1}{3}b_m b^m + \frac{i}{3}b_m(K_{i^*}\partial^m\bar{A}_i-K_i\partial^m A_i)+ \frac{2}{3}J'b_m B^m+ J'D + K_{ij^*}F_i\overbar{F}_j\\-\bigg(\frac{1}{3}{J'}^2-\frac{1}{2}J''\bigg)X\overbar{X}-\frac{1}{3}(M\overbar{M}+K_i K_{j^*}F_i\overbar{F}_j-J'K_{i^*}\overbar{F}_i X-J'K_i F_i\overbar{X}+K_{i^*}\overbar{F}_i\overbar{M}+K_iF_iM-J'MX-J'\overbar{M}\overbar{X}) \biggl\}\\+\frac{1}{2}D^2+F_i\mathcal{W}_i+\overbar{F}_i\overbar{\mathcal{W}}_i-\overbar{M}\mathcal{W}-M\overbar{\mathcal{W}}~.
\end{multline}
In our equations above, the $K$, $J$ and $\mathcal{W}$ now represent the lowest components of the corresponding  superfields, being functions of the scalar fields $A_i$ and $C$. As regards their derivatives, we have used the notation  $K_i\equiv\frac{\partial K}{\partial A_i}$, $K_{i^*}\equiv\frac{\partial K}{\partial\overbar{A}_i}$, $K_{ij^*}\equiv\frac{\partial^2K}{\partial A_i\partial\overbar{A}_j}$, $J'\equiv\frac{\partial J}{\partial C}$, $\mathcal{W}_i\equiv\frac{\partial\mathcal{W}}{\partial A_i}$, $\overbar{\mathcal{W}}_i\equiv\frac{\partial\overbar{\mathcal{W}}}{\partial\overbar{A}_i}$.

In order to eliminate the auxiliary fields in accordance to their algebraic equations of motion,
we first separate $M$, $F_i$ and $X$ from each other via a substitution,
\begin{align}
&M=N+J'\overbar{X}-K_{i^*}\overbar{F}_i~,\\
&\overbar{M}=\overbar{N}+J'X-K_iF_i~.
\end{align}
In terms of the new auxiliary fields $N$ and $\overbar{N}$, the auxiliary part of the Lagrangian takes the form
\begin{multline}
e^{-1}\mathcal{L}_{\rm aux.}=e^{-\frac{1}{3}(K+2J)}\biggl\{\frac{1}{3}b_m b^m + \frac{i}{3}b_m(K_{i^*}\partial^m\bar{A}_i-K_i\partial^m A_i)+ \frac{2}{3}J'b_m B^m+ J'D + K_{ij^*}F_i\overbar{F}_j\\+\frac{1}{2}J''X\overbar{X}-\frac{1}{3}N\overbar{N} \biggl\}+\frac{1}{2}D^2+F_i\mathcal{W}_i+\overbar{F}_i\overbar{\mathcal{W}}_i-\mathcal{W}(\overbar{N}+J'X-K_iF_i)-\overbar{\mathcal{W}}(N+J'\overbar{X}-K_{i^*}\overbar{F}_i)~,
\end{multline}
so that Euler-Lagrange equations of the auxiliary fields are easily solved as
\begin{gather*}
b_m=-J'B_m-\frac{i}{2}(K_{i^*}\partial_m\bar{A}_i-K_i\partial_m A_i)~,\\
D=-J'e^{-\frac{1}{3}(K+2J)},\;\;\;\; N=-3e^{\frac{1}{3}(K+2J)}\mathcal{W}~,\\
F_i=-e^{\frac{1}{3}(K+2J)}K_{ij^*}^{-1}(\overbar{\mathcal{W}}_j+K_{j^*}\overbar{\mathcal{W}}),\;\;\;\; X=2\frac{J'}{J''}e^{\frac{1}{3}(K+2J)}\overbar{\mathcal{W}}~.
\end{gather*}

After a substitution of those solutions back into the Lagrangian, we find
\begin{multline}
e^{-1}\mathcal{L}=e^{-\frac{1}{3}(K+2J)}\biggl\{
	-\frac{1}{2}R-K_{ij*}\partial_m A_i\partial^m\bar{A}_j-\frac{1}{6}K_iK_{j^*}\partial_mA_i\partial^m\bar{A}_j-\frac{1}{12}K_iK_j\partial_mA_i\partial^mA_j\\-\frac{1}{12}K_{i^*}K_{j^*}\partial_m\bar{A}_i\partial^m\bar{A}_j-\bigg(\fracmm{1}{3}{J'}^2-\fracmm{1}{2}J''\bigg)\partial_m C\partial^m C+J'\Box C-\frac{1}{3}J'\partial_mC(K_{i^*}\partial^m\bar{A}_i+K_i\partial^mA_i)-\frac{1}{2}J''B_mB^m
	\biggl\}\\
-\frac{1}{4}F_{mn}F^{mn}-\frac{1}{2}e^{-\frac{2}{3}(K+2J)}{J'}^2-e^{\frac{1}{3}(K+2J)}\biggl[
	K^{-1}_{ij^*}(\mathcal{W}_i+K_i\mathcal{W})(\overbar{\mathcal{W}}_j+K_{j^*}\overbar{\mathcal{W}})-\bigg(3-2\fracmm{{J'}^2}{J''}\bigg)\mathcal{W}\overbar{\mathcal{W}}~.
	\biggl]
\end{multline}

A transition from Jordan to Einstein frame is achieved by Weyl rescaling of spacetime metric,
\begin{gather*}
g_{mn}\rightarrow e^{\Lambda}g_{mn}~,\;\;\;\; e\rightarrow e^{2\Lambda}e~,\quad {\rm with}\quad
\Lambda=\frac{1}{3}(K+2J)~.
\end{gather*}
Then the scalar curvature term transforms as
\begin{equation}
-\frac{1}{2}ee^{-\frac{1}{3}(K+2J)}R\rightarrow-\frac{1}{2}eR+\frac{1}{12}e(\partial_mK+2\partial_mJ)^2~.
\end{equation} 

It gives rise to the Lagrangian
\begin{equation} \label{complag}
e^{-1}\mathcal{L}=-\frac{1}{2}R-K_{ij*}\partial_m A_i\partial^m\bar{A}_j-\frac{1}{4}F_{mn}F^{mn}-\frac{1}{2}J''\partial_mC\partial^mC-\frac{1}{2}J''B_mB^m-\mathcal{V},
\end{equation}
with the scalar potential
\begin{equation} \label{pot}
\mathcal{V}=\frac{1}{2}{J'}^2+e^{K+2J}{}\biggl[
	K^{-1}_{ij^*}(\mathcal{W}_i+K_i\mathcal{W})(\overbar{\mathcal{W}}_j+K_{j^*}\overbar{\mathcal{W}})-\bigg(3-2\fracmm{{J'}^2}{J''}\bigg)\mathcal{W}\overbar{\mathcal{W}}
	\biggl].
\end{equation}

Equations (\ref{sslag}), (\ref{complag}) and (\ref{pot}) are our main results in this Section. When the real superfield 
$V$ is dropped $(J=0)$, our result coincides with the standard Lagrangian and the scalar potential of chiral superfields in $N=1$ supergravity \cite{Cremmer:1978hn}. When all the chiral superfields $\Phi_i$ are dropped 
$(K=\mathcal{W}=0)$, our results coincide with
those in \cite{Ferrara:2013rsa,VanProeyen:1979ks}.~\footnote{Our notation for $J$ differs by the sign from that of
 \cite{Ferrara:2013rsa,VanProeyen:1979ks}.}

As is clear from (\ref{complag}), the absence of ghosts requires $J''(C)>0$.

\section{Vacuum solution}

In this Section we restrict ourselves to a single chiral superfield $\Phi$ having the canonical K\"ahler potential and the superpotential given by a sum of a linear term and a constant,
\begin{equation} \label{polonyi}
K= \Phi\overbar{\Phi}~,\qquad \mathcal{W}=\mu(\Phi +\beta)~.
\end{equation}
This particular choice is known in the literature as {\it Polonyi model} \cite{Polonyi:1977pj}.~\footnote{It is worth mentioning that this choice is most natural for a nilpotent (Akulov-Volkov) superfield, $\Phi^2=0$.}

In accordance to the previous Section, it gives rise to the Lagrangian
\begin{multline}
e^{-1}\mathcal{L}=-\frac{1}{2}R-\partial_mA\partial^m\bar{A}-\frac{1}{4}F_{mn}F^{mn}-\frac{1}{2}J''\partial_mC\partial^mC-\frac{1}{2}J''B_mB^m-\frac{1}{2}{J'}^2\\-\mu^2e^{A\bar{A}+2J}\biggl[
	|1+A\beta+A\bar{A}|^2-\bigg(3-2\fracmm{{J'}^2}{J''}\bigg)|A+\beta|^2
	\biggl]~.
\end{multline}

The (Minkowski) vacuum conditions in this model are given by 
\begin{gather}
V=\frac{1}{2}{J'}^2+\mu^2e^{A\bar{A}+2J}\biggl[
	|1+A\beta+A\bar{A}|^2-\bigg(3-2\fracmm{{J'}^2}{J''}\bigg)|A+\beta|^2
	\biggl]=0~,\\
\partial_{\bar{A}}V=A\tilde{V}_F+\mu^2e^{A\bar{A}+2J}\biggl[
A(1+\bar{A}\beta+A\bar{A})+(A+\beta)(1+A\beta+A\bar{A})-\bigg(3-2\fracmm{{J'}^2}{J''}\bigg)(A+\beta)
\biggl]=0~,\\
\partial_C V=J'\biggl\{J''+2\mu^2e^{A\bar{A}+2J}\biggl[
	|1+A\beta+A\bar{A}|^2-\bigg(1-2\fracmm{{J'}^2}{J''}+\fracmm{J'J'''}{{J''}^2}\bigg)|A+\beta|^2
	\biggl]\biggl\}=0~,
\end{gather}
where we have introduced $\tilde{V}_F$ as the F-type scalar potential with the additional $J$-dependent term as
\begin{equation}
\tilde{V}_F=\mu^2e^{A\bar{A}+2J}\biggl[
	|1+A\beta+A\bar{A}|^2-\bigg(3-2\fracmm{{J'}^2}{J''}\bigg)|A+\beta|^2
	\biggl].
\end{equation}

A simple solution to those equations exist when $J'=0$, which separates the Polonyi multiplet from the vector multiplet. The remaining vacuum equations allow a solution with the VEV $\langle A\rangle\equiv\alpha=(\sqrt{3}-1)$ and $\beta=2-\sqrt{3}$ \cite{Polonyi:1977pj}. This celebrated (Polonyi) solution describes a stable Minkowski vacuum with spontaneously broken SUSY since $\langle F\rangle=\mu$. Hence, the parameter $\mu$ defines the scale of SUSY breaking, which is {\it arbitrary} in this model. The related gravitino mass is given by 
$m_{3/2}=\mu e^{2-\sqrt{3}+\langle J\rangle}$. There is also a massive scalar of mass $2m_{3/2}$ and a massless fermion in the
physical spectrum. 

It should be emphasized that the Polonyi field does not affect inflation associated with the scalar $C$ as the inflaton
belonging to the massive vector multiplet, and having the $D$-type scalar potential $V(C)=\frac{1}{2}{J'}^2$ with arbitrary real $J$-function.  Of course, the true inflaton field should be canonically normalized via the appropriate field redefinition of $C$.

When trying to get other patterns of SUSY breaking after inflation by demanding $J'\neq 0$ and $\alpha=\beta=0$,  we get  two conditions on the $J$-function,
\begin{gather}
{J'}^2=J''\label{jp}~,\\
J''=-2\mu^2e^{2J}~.\label{jp2}
\end{gather}
The first equation is solved by $J=-\log C+const.$, then the second condition yields the consistency 
relation $const.=-\frac{1}{2}\log(-2\mu^2)$. Since both $J$ and $\mu$ should be real, there is no solution.
However, when allowing $\beta\neq 0$, the second equation \eqref{jp2} gets modified as 
\begin{equation}
J''=-2\mu^2e^{2J}(1-\beta^2)~,\label{jp3}
\end{equation}
so that the reality of $J$ and $\mu$ requires $\beta>1$. Then \eqref{jp} reads ${J'}^2=C^{-2}$ and is easily solvable.
However, such scalar potential is not suitable for inflation (no slow roll). More general vacuum solutions with 
 $J'\neq 0$ will be investigated elsewhere.

\newpage 

\section{Conclusion}

Our basic equations (\ref{sslag}), (\ref{complag}) and (\ref{pot}) supply new theoretical tools for the
inflationary model building in supergravity. They can be further generalized e.g., by including an extra
function $g(\Phi)$ of the chiral superfields in front of the vector multiplet kinetic term in (\ref{sslag}), and/or
replacing the Maxwell-type kinetic term of the vector multiplet by the Born-Infeld-type action, like e.g., in 
 \cite{Abe:2015fha}.

Our models have three arbitrary (input) potentials $K$, $\mathcal{W}$ and $J$, providing more flexibility to the inflationary model building and, perhaps,  being derivable from a more fundamental theory, like string theory.

Our construction does not have an R-symmetry, and is apparently unrelated to (the dual version of) matter-coupled $(R+R^2)$ supergravity in its 'new-minimal' formulation \cite{extra}.

In particular, as was demonstrated in Sec.~3, our construction easily supplies spontaneous SUSY breaking after inflation to the supergravity-based inflationary models whose inflaton belongs to a massive vector multiplet, via  their coupling to Polonyi multiplet. Those models are {\it limited} in the sense that they provide the {\it minimal} extension of the inflationary models proposed in \cite{Ferrara:2013rsa} for the sake of spontaneous SUSY breaking in Minkowski vacuum after inflation. The Polonyi multiplet itself can be assigned to the hidden sector needed for SUSY breaking 
and its gravitational mediation to the visible sector in more general field-theoretical models of particle physics beyond the Standard Model in the context of supergravity.

\section*{Acknowledgements}

YA is supported by a scholarship from the Ministry of Education, Culture, Sports, Science and Technology (MEXT) in Japan.
SVK is grateful to I. Antoniadis, A. Chatrabhuti and O. Evnin for discussions. SVK is supported by a Grant-in-Aid of the Japanese Society for Promotion of Science (JSPS) under No.~2640025200, a TMU President Grant of Tokyo Metropolitan University in Japan, the World Premier International Research Center Initiative (WPI Initiative), MEXT, Japan, the CUniverse research promotion project by Chulalongkorn University (grant reference CUAASC) in Bangkok, Thailand, and the Competitiveness Enhancement Program of Tomsk Polytechnic University in Russia.

\bibliographystyle{utphys} 
%\bibliography{bibliography.bib}

\begin{thebibliography}{10}


\bibitem{Ade:2015xua}
{\bfseries Planck} Collaboration, P.~A.~R. Ade {\em et~al.}, ``{Planck 2015
  results. XIII. Cosmological parameters},''
\href{http://arxiv.org/abs/1502.01589}{{\ttfamily arXiv:1502.01589
  [astro-ph.CO]}}.
%%CITATION = ARXIV:1502.01589;%%.

\bibitem{Ade:2015lrj}
{\bfseries Planck} Collaboration, P.~A.~R. Ade {\em et~al.}, ``{Planck 2015
  results. XX. Constraints on inflation},''
\href{http://arxiv.org/abs/1502.02114}{{\ttfamily arXiv:1502.02114
  [astro-ph.CO]}}.
%%CITATION = ARXIV:1502.02114;%%.

\bibitem{Array:2015xqh}
{\bfseries BICEP2, Keck Array} Collaboration, P.~A.~R. Ade {\em et~al.},
  ``{Improved Constraints on Cosmology and Foregrounds from BICEP2 and Keck
  Array Cosmic Microwave Background Data with Inclusion of 95 GHz Band},''
  \href{http://dx.doi.org/10.1103/PhysRevLett.116.031302}{{\em Phys. Rev.
  Lett.} {\bfseries 116} (2016) 031302},
\href{http://arxiv.org/abs/1510.09217}{{\ttfamily arXiv:1510.09217
  [astro-ph.CO]}}.
%%CITATION = ARXIV:1510.09217;%%.

\bibitem{Yamaguchi:2011kg}
M.~Yamaguchi, ``{Supergravity based inflation models: a review},''
  \href{http://dx.doi.org/10.1088/0264-9381/28/10/103001}{{\em Class. Quant.
  Grav.} {\bfseries 28} (2011) 103001},
\href{http://arxiv.org/abs/1101.2488}{{\ttfamily arXiv:1101.2488
  [astro-ph.CO]}}.
%%CITATION = ARXIV:1101.2488;%%.

\bibitem{Ketov:2012yz}
S.~V. Ketov, ``{Supergravity and Early Universe: the Meeting Point of Cosmology
  and High-Energy Physics},''
  \href{http://dx.doi.org/10.1142/S0217751X13300214}{{\em Int. J. Mod. Phys.}
  {\bfseries A28} (2013) 1330021},
\href{http://arxiv.org/abs/1201.2239}{{\ttfamily arXiv:1201.2239 [hep-th]}}.
%%CITATION = ARXIV:1201.2239;%%.

\bibitem{Goncharov:1983mw}
A.~B. Goncharov and A.~D. Linde, ``{Chaotic Inflation in Supergravity},''
\href{http://dx.doi.org/10.1016/0370-2693(84)90027-3}{{\em Phys. Lett.}
  {\bfseries B139} (1984) 27--30}.
%%CITATION = PHLTA,B139,27;%%.

\bibitem{Kawasaki:2000yn}
M.~Kawasaki, M.~Yamaguchi, and T.~Yanagida, ``{Natural chaotic inflation in
  supergravity},'' \href{http://dx.doi.org/10.1103/PhysRevLett.85.3572}{{\em
  Phys. Rev. Lett.} {\bfseries 85} (2000) 3572--3575},
\href{http://arxiv.org/abs/hep-ph/0004243}{{\ttfamily arXiv:hep-ph/0004243
  [hep-ph]}}.
%%CITATION = HEP-PH/0004243;%%.

\bibitem{Kallosh:2010ug}
R.~Kallosh and A.~Linde, ``{New models of chaotic inflation in supergravity},''
  \href{http://dx.doi.org/10.1088/1475-7516/2010/11/011}{{\em JCAP} {\bfseries
  1011} (2010) 011},
\href{http://arxiv.org/abs/1008.3375}{{\ttfamily arXiv:1008.3375 [hep-th]}}.
%%CITATION = ARXIV:1008.3375;%%.

\bibitem{Kallosh:2010xz}
R.~Kallosh, A.~Linde, and T.~Rube, ``{General inflaton potentials in
  supergravity},'' \href{http://dx.doi.org/10.1103/PhysRevD.83.043507}{{\em
  Phys. Rev.} {\bfseries D83} (2011) 043507},
\href{http://arxiv.org/abs/1011.5945}{{\ttfamily arXiv:1011.5945 [hep-th]}}.
%%CITATION = ARXIV:1011.5945;%%.

\bibitem{Abe:2014opa}
H.~Abe, S.~Aoki, F.~Hasegawa, and Y.~Yamada, ``{Illustrating SUSY breaking
  effects on various inflation mechanisms},''
  \href{http://dx.doi.org/10.1007/JHEP01(2015)026}{{\em JHEP} {\bfseries 01}
  (2015) 026},
\href{http://arxiv.org/abs/1408.4875}{{\ttfamily arXiv:1408.4875 [hep-th]}}.
%%CITATION = ARXIV:1408.4875;%%.

\bibitem{Ketov:2014qha}
S.~V. Ketov and T.~Terada, ``{Inflation in supergravity with a single chiral
  superfield},'' \href{http://dx.doi.org/10.1016/j.physletb.2014.07.036}{{\em
  Phys. Lett.} {\bfseries B736} (2014) 272--277},
\href{http://arxiv.org/abs/1406.0252}{{\ttfamily arXiv:1406.0252 [hep-th]}}.
%%CITATION = ARXIV:1406.0252;%%.

\bibitem{Ketov:2014hya}
S.~V. Ketov and T.~Terada, ``{Generic Scalar Potentials for Inflation in
  Supergravity with a Single Chiral Superfield},''
  \href{http://dx.doi.org/10.1007/JHEP12(2014)062}{{\em JHEP} {\bfseries 12}
  (2014) 062},
\href{http://arxiv.org/abs/1408.6524}{{\ttfamily arXiv:1408.6524 [hep-th]}}.
%%CITATION = ARXIV:1408.6524;%%.

\bibitem{Ketov:2016gej}
S.~V. Ketov and T.~Terada, ``{On SUSY Restoration in Single-Superfield
  Inflationary Models of Supergravity},''
\href{http://arxiv.org/abs/1606.02817}{{\ttfamily arXiv:1606.02817 [hep-th]}}.
%%CITATION = ARXIV:1606.02817;%%.

\bibitem{Ferrara:2014kva}
S.~Ferrara, R.~Kallosh, and A.~Linde, ``{Cosmology with Nilpotent
  Superfields},'' \href{http://dx.doi.org/10.1007/JHEP10(2014)143}{{\em JHEP}
  {\bfseries 10} (2014) 143},
\href{http://arxiv.org/abs/1408.4096}{{\ttfamily arXiv:1408.4096 [hep-th]}}.
%%CITATION = ARXIV:1408.4096;%%.

\bibitem{Ferrara:2013rsa}
S.~Ferrara, R.~Kallosh, A.~Linde, and M.~Porrati, ``{Minimal Supergravity
  Models of Inflation},''
  \href{http://dx.doi.org/10.1103/PhysRevD.88.085038}{{\em Phys. Rev.}
  {\bfseries D88} no.~8, (2013) 085038},
\href{http://arxiv.org/abs/1307.7696}{{\ttfamily arXiv:1307.7696 [hep-th]}}.
%%CITATION = ARXIV:1307.7696;%%.

\bibitem{VanProeyen:1979ks}
A.~Van~Proeyen, ``{Massive Vector Multiplets in Supergravity},''
\href{http://dx.doi.org/10.1016/0550-3213(80)90345-4}{{\em Nucl. Phys.}
  {\bfseries B162} (1980) 376}.
%%CITATION = NUPHA,B162,376;%%.

\bibitem{Polonyi:1977pj}
J.~Polonyi,
``{Generalization of the Massive Scalar Multiplet Coupling to the
  Supergravity},''.
%%CITATION = KFKI-77-93;%%.

\bibitem{Wess:1992cp}
J.~Wess and J.~Bagger, {\em {Supersymmetry and supergravity}}.
\newblock
1992.
\newblock
%%CITATION = INSPIRE-350988;%%.

\bibitem{Cremmer:1978hn}
E.~Cremmer, B.~Julia, J.~Scherk, S.~Ferrara, L.~Girardello, and P.~van
  Nieuwenhuizen, ``{Spontaneous Symmetry Breaking and Higgs Effect in
  Supergravity Without Cosmological Constant},''
\href{http://dx.doi.org/10.1016/0550-3213(79)90417-6}{{\em Nucl. Phys.}
  {\bfseries B147} (1979) 105}.
%%CITATION = NUPHA,B147,105;%%.

\bibitem{Abe:2015fha}
H.~Abe, Y.~Sakamura, and Y.~Yamada, ``{Massive vector multiplet inflation with
  Dirac-Born-Infeld type action},''
  \href{http://dx.doi.org/10.1103/PhysRevD.91.125042}{{\em Phys. Rev.}
  {\bfseries D91} no.~12, (2015) 125042},
\href{http://arxiv.org/abs/1505.02235}{{\ttfamily arXiv:1505.02235 [hep-th]}}.
%%CITATION = ARXIV:1505.02235;%%.

\bibitem{extra} 
S.~Ferrara and M.~Porrati, ``{Minimal $R+R^2$ supergravity models of inflation coupled to matter},''
\href{http://dx.doi.org/10.1016/j.physletb.2014.08.050}{{\em Phys. Lett.}
{\bfseries B737} (2014) 135-138},
\href{http://arxiv.org/abs/1407.6164}{{\ttfamily arXiv:1407.6164 [hep-th]}}.
%% CITATION = ARXIV:1407.6164;%%


\end{thebibliography}

\providecommand{\href}[2]{#2}\begingroup\raggedright
\endgroup

\end{document}